\documentclass[reprint,pra,amsmath,amssymb,aps,showpacs]{revtex4-1}
\usepackage{graphicx}% Include figure files
\usepackage{dcolumn}% Align table columns on decimal point
\usepackage{bm}% bold math
\usepackage[colorlinks,linkcolor=blue,citecolor=blue]{hyperref} 
\begin{document}

\preprint{APS/123-QED}

\title{Action differences between fixed points and accurate heteroclinic orbits}

\author{Jizhou Li}
\author{Steven Tomsovic}
\affiliation{Department of Physics and Astronomy, Washington State University, Pullman, WA, USA 99164-2814}

\date{\today}

\begin{abstract}

A general relation is derived for the action difference between two fixed points and a phase space area bounded by the irreducible component of a heteroclinic tangle.  The determination of this area can require accurate calculation of heteroclinic orbits, which are important in a wide range of dynamical system problems.  For very strongly chaotic systems initial deviations from a true orbit are magnified by a large exponential rate making direct computational methods fail quickly.  Here, a method is developed that avoids direct calculation of the orbit by making use of the well-known stability property of the invariant unstable and stable manifolds. Under an area-preserving map, this property assures that any initial deviation from the stable (unstable) manifold collapses onto themselves under inverse (forward) iterations of the map.  Using a set of judiciously chosen auxiliary points on the manifolds, long orbit segments can be calculated using the stable and unstable manifold intersections of the heteroclinic (homoclinic) tangle.  Detailed calculations using the example of the kicked rotor are provided along with verification of the relation between action differences.  The loop structure of the heteroclinic tangle is necessarily quite different from that of the turnstile for a homoclinic tangle, its analogous partner.

\end{abstract}

    % Computational methods in classical mechanics
							  % Low-dimensional chaos
					 		  % Numerical simulations of chaotic systems

\maketitle

\section{Introduction}
\label{Intro}

There are significant classes of problems in physics and chemistry for which it is very important to have accurate information about periodic, heteroclinic, and/or homoclinic trajectories in chaotic dynamical systems.  For example, they arise in semiclassical trace formulae~\cite{Gutzwillerbook} and wave packet propagation~\cite{Tomsovic91b,Tomsovic93}.  Trace formulae have been invoked to understand shell structure in nuclei,  metallic clusters, and quantum dots~\cite{Brack06}, and the Bohigas-Gianonni-Schmit conjecture relating chaotic systems quantum spectra to random matrix theory~\cite{Heusler07}.  The time evolution of wave packets have been used for understanding driven cold atoms~\cite{Bakman15}, electrons in strong fields~\cite{Zagoya14}, fidelity studies~\cite{Jalabert01,Cerruti02}, and a broad range of spectroscopic and pump-probe experiments~\cite{Heller81b,Alber91}.  If a particular system of interest possesses a strongly chaotic dynamics, then a semiclassical approximation leads to sums over finite-time segments of such trajectories~\cite{Tomsovic93}.  Furthermore, the quantum mechanical phases are largely controlled by Hamilton's Principle Function (or classical action for short) for these segments divided by Planck's constant.  As a result, deep in a semiclassical regime, small changes in classical actions or small action differences between the various trajectory segments may result in significant changes in interferences quantum mechanically.   The actions must be known accurately to predict these interferences correctly.

As time increases in the semiclassical sums, the number of contributing terms increases exponentially rapidly, but there cannot be an exponentially increasing amount of information present in the quantum propagation.  This situation has to be reflected in the classical dynamics through the existence of correlations in classical actions~\cite{Sieber01}.  A critical element that bears on the correlations is the relationship between the limiting differences of certain heteroclinic trajectory pairs and areas enclosed in phase space~\cite{MacKay87}.  These areas show up in multiple action differences and their boundaries are determined by the heteroclinic or homoclinic tangles themselves.  Our purpose is to relate periodic orbit or fixed point action differences to phase space areas defined by heteroclinic tangles, and develop a scheme in which heteroclinic orbits of strongly chaotic systems are calculated accurately enough and over a long enough time interval that the limiting classical action differences give precise evaluations of such critical areas.  

It turns out that for a chaotic system with a large enough Lyapunov exponent, using Hamilton's equations with any phase point on some particular orbit (or if it is a mapping that is of interest, the map) will fail to faithfully follow a heteroclinic orbit segment, except for very short segments.  In order to do better than this, an alternative computational scheme is desirable.  Previous papers on this topic were mainly focused on continuous time systems, where the infinite time interval associated with the homoclinic or heteroclinic orbits are truncated into finite time domains, with appropriate boundary value conditions so that the standard boundary value problem solvers apply~\cite{Doedel86,Doedel89,Beyn90,Bai93}.  Modified approaches using the arclength of the orbit as the system parameter instead of time were given in~\cite{Moore95b} and~\cite{Liu97}, which avoid the truncation process but still require solving the boundary value problem using numerical discretization and collocation techniques.  Methods using spectral expansions that also avoid the finite truncation were given in~\cite{Liu94}.  Recent papers facilitating other numerical techniques can be found in~\cite{Korostyshevskiy07}, where Hermite-Fourier expansions were used to approximate the homoclinic solutions; and in~\cite{Dong14}, where a variational approach similar to the one proposed in~\cite{Lan04} was employed.  The method developed here is vastly simpler than those mentioned above.  It focuses on Hamiltonian systems with invariant manifolds, and relies on the fact that a heteroclinic (homoclinic) orbit lies at all times at the intersection of a stable and unstable manifold.  These manifolds can be computed in quite stable ways, and therefore their intersections can also be calculated in just as stable ways.  It is conceptually straightforward, also extremely fast in terms of calculation time.  No special numerical techniques such as collocations in the boundary value problems or algebraic equation solvers are needed.  Therefore, the degree of precision is only affected by the density of data points interpolating the manifolds.  If the set of data points are dense enough, which can be easily arranged by inserting more initial points for generating the manifolds, the precision is only limited by the interpolation technique used to intersect the stable and unstable manifolds, and machine precision, whichever is greater.  

By switching initial intersections (denoted by $R_0$ ahead), our method can be used to find any heteroclinic (homoclinic) orbit with arbitrarily long excursion length, not being restricted to some particular orbit.  Thus, an enumeration of the infinite set of orbits is made possible by switching the initial intersections, $R_0$, and repeated use of this method.  In general, there is a minimal or irreducible structure from which to select an $R_0$ in order to obtain any orbit of one's choosing and to avoid duplications.  For the homoclinic case, this is the well known turnstile (assuming no special symmetries), and for the heteroclinic case a necessarily different loop structure.  The relation between phase space areas and periodic orbit action differences reinforces this.  Without loss of generality, from here on, we concentrate on dynamical maps as a continuous dynamical system can be reduced to a Poincar\'e map.  Thus, a heteroclinic orbit segment is constructed as a sequence of manifold intersections forward or inverse in time (iteration number) from some given intersection phase point as opposed to the forward or inverse mapping of that same phase point.

This paper is organized as follows, the next section covers necessary background and the introduction of some useful notation.  All of the main ideas are illustrated using the kicked rotor with a strong kicking strength, a well-known, simple paradigm of strongly chaotic dynamics.  The following section describes the use of the well known structural stability inherent in chaotic dynamical systems~\cite{Ott02} in order to construct stable and unstable manifolds.  This is followed by a technique to locate intersecting points. The technique is used to compare stability exponents of certain fixed points of the map calculated using the stability matrix and the constructed orbit segments, and to compare certain phase space areas with limiting classical action differences.  The area of the heteroclinic loop structure is shown to be the action difference of the fixed points.  Finally, we summarize and note some ideas of interest for future work.

 \section{Background}
 \label{background}
 
 Let $X=(q,p)$ represent points in phase space and classical transport be described by a time-dependent correlation function between an initially localized Gaussian density of phase space points $\rho_{\alpha}$ centered at point $X_\alpha$ and a final destination Gaussian density $\rho_{\beta}$ centered at point $X_\beta$.  In the limit of highly localized densities, it can be expressed as a sum over all heteroclinic trajectory segments~\cite{Tomsovic93}
\begin{equation}\label{eq:one}
\Gamma_{\beta \alpha}(t) \longrightarrow \sum_{\gamma}\big(\rho_{\beta},T_\gamma^t \rho_{\alpha}\big)
\end{equation}
where $T$ is a linearized dynamical mapping of points $X$, $t$ is the time, i.e.~number of iterations of the map, and $\gamma$ denotes those segments that take time $t$ to leave the neighborhood of $X_\alpha$ and arrive in the neighborhood of $X_\beta$ (they serve as the orbit segments about which the linearizations are done).  For simplicity, let $X_\alpha$ and $X_\beta$ be fixed points of the map $T$.  The segments belong to heteroclinic orbits that converge to $X_\alpha$ for $t\to -\infty$ and converge to $X_\beta$ for $t\to +\infty$.  The quantum mechanical analog in the semiclassical limit of the correlation function is~\cite{Tomsovic93}
\begin{equation}\label{eq:qm}
{\cal C}_{\beta \alpha}(t) \approx \sum_{\gamma}\left\langle \beta|U_\gamma(t)| \alpha\right\rangle
\end{equation}
where $|\alpha\rangle$ is the ket vector corresponding to a quantum wave packet centered at $X_\alpha$ and $U_\gamma (t)$ is an appropriately linearized unitary time translation operator.  The summation is over the same heteroclinic orbit segments.  These two equations hold equally well for open or bounded systems.  The quantum expression is our main motivation for investigating the heteroclinic orbits and their properties.

For a dynamical map a complete orbit $\lbrace R_{0}\rbrace$ can be represented by a bi-infinite sequence of the form: $\lbrace R_{-\infty}, \dots, R_{-2}, R_{-1}, R_{0}, R_{1}, R_{2}, \dots, R_{\infty}\rbrace$ where $R_{j}$ maps to the point on the orbit $R_{j+t}$ after $t$ iterations of the map.  The times have a translational arbitrariness to them, but here they are set such that $0$ is the time for a presumed known initial condition that defines a particular orbit.  Given that in calculations the mapping itself cannot be applied exactly and repeated mappings lead to exponentially growing errors, a true orbit cannot be found this way.  Ahead, we will choose a case for illustration whose dynamics are so unstable (large Lyapunov exponent) that it is not possible to follow a heteroclinic orbit segment more than $\pm 5$ iterations via the mapping (using double precision).  Because our interest is in the set of heteroclinic orbits, it is not possible to rely on the Shadowing Lemma.  While there is a true orbit of the system "shadowing" an orbit found with the mapping, the likelihood that it is the actual heteroclinic orbit of interest is vanishingly small.

\subsection{The kicked rotor}
\label{kickedrotor}

The kicked rotor on a torus has been a simple, yet extraordinary paradigm for chaotic systems for roughly 50 years and a great deal is known about it~\cite{Chirikov79}.  It is a mechanical-type particle constrained to
move on a ring that is kicked instantaneously every multiple of a unit
time, $t=n$.  The Hamiltonian takes the form 
\begin{equation}
\label{krg}
H(q,p) = \frac{p^2}{2} - \frac{K}{4\pi^2}\cos 2\pi q \sum_{n=-\infty}^\infty \delta(t-n) 
\end{equation}
The mapping equations are:
\begin{equation} \label{eq:two}
\begin{split}
& p_{n+1} =p_{n}-\frac{K}{2\pi }\sin 2\pi q_{n} \pmod 1 \\
& q_{n+1} =q_{n}+p_{n+1} \pmod 1
\end{split}
\end{equation} 
As the kicking strength parameter $K$ increases away from zero, the system becomes more and more chaotic.  For $K$ values exceeding roughly $2\pi$, the system is very nearly completely and strongly chaotic.  The Lyapunov exponent $\lambda$ is known analytically to be~\cite{Tomsovic07}
\begin{equation}
\lambda \sim \ln \frac{K}{2} + \frac{1}{K^2-4} + O\left(\frac{1}{\left(K^2-4\right)^3}\right)
\end{equation}
In all the calculations ahead, the kicking strength is taken as $K=8.25$, which is a strongly chaotic case.  In addition, there are two very convenient fixed points of the mapping, $(0,0)$ and $(0.5,0)$, to be used for $X_\alpha$ and $X_\beta$ respectively.  Therefore, the heteroclinic orbits discussed ahead lie at intersections of the unstable manifold of the phase point $(0,0)$ with the stable manifold of the phase point $(0.5,0)$.

It turns out to be convenient to use the ``unfolded torus", meaning that by not invoking the modulus $1$ operations in the mapping equations, there is a ``flat" phase space that extends to infinity.  Each unit square is a repetition of the fundamental torus which is the $[0,1) \times [0,1)$ square in the phase space.  Any two points that are separated by integer numbers on either the $q$ or $p$ coordinates are the same point; i.e.~$(q,p)$ and $(q+n_{q},p+n_{p})$ are the same point if both $n_{q}$ and $n_{p}$ are integers.  The integers $n_{q}$ and $n_{p}$ can be thought of as winding numbers (including negative integers), i.e.~how many times a particle has wrapped around the cycles of the torus, which can have phase consequences in quantum mechanics.

\begin{figure}[h]
\centering
{\includegraphics[width=8cm]{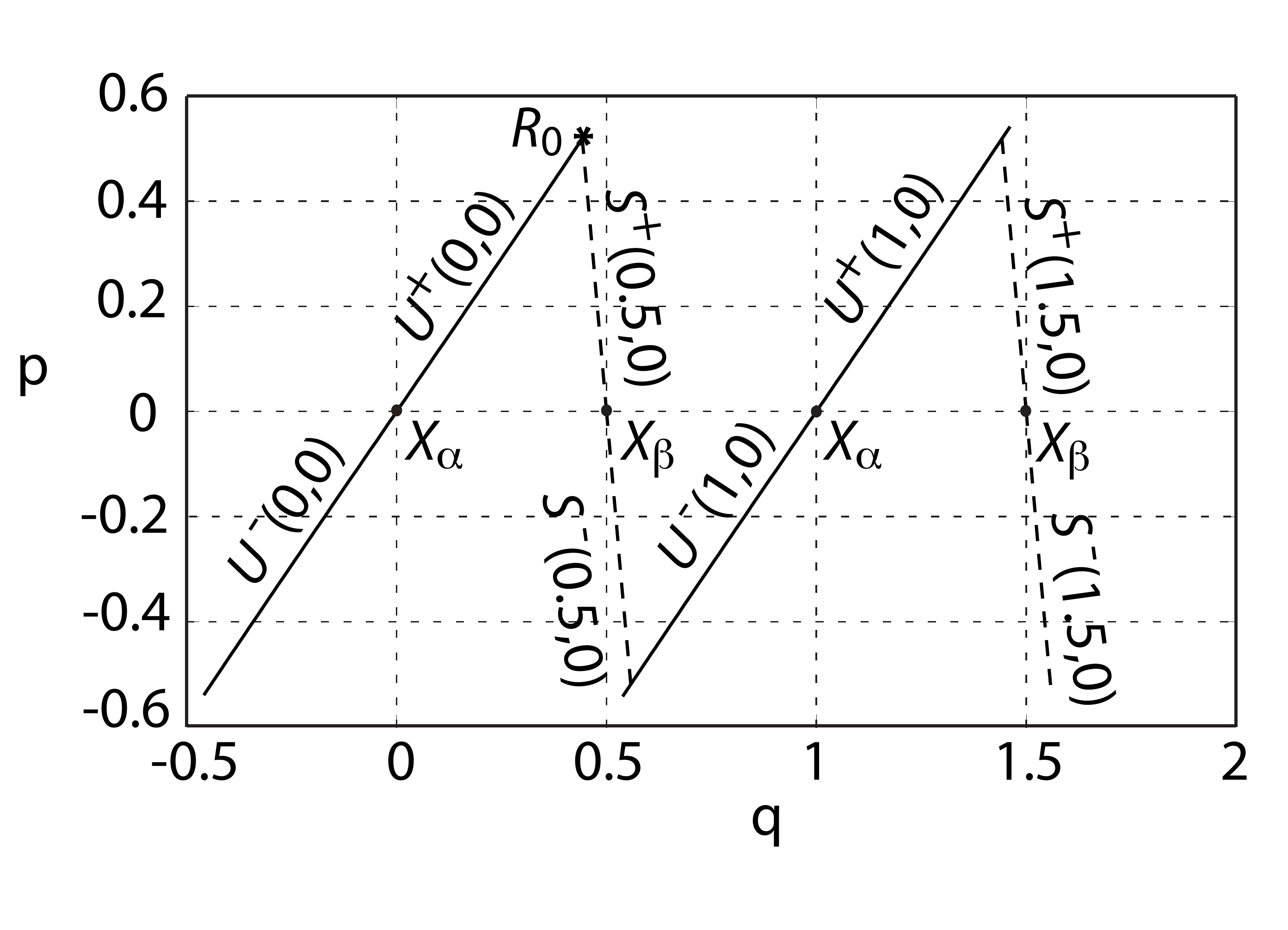}}
\caption{The initial segments of the unstable manifolds (shown in solid curves) of (0,0) and (1,0), and the stable manifolds (shown in dashed curves) of (0.5,0) and (1.5,0) on the unfolded torus. $R_{0}$ is a heteroclinic intersection point between $U^{+}(0,0)$ and $S^{+}(0.5,0)$.  For all the figures in this paper, the unstable manifold is plotted as solid curves and the stable manifolds in dashed curves.}
\label{fig:one}
\end{figure}

As shown in Fig.~\ref{fig:one}, we define the part of the unstable manifold (solid curves) of $X_\alpha$ which is initially pointing to the upper-right direction to be the upper branch of the manifold, denoting it by $U^{+}(0,0)$. Denote the piece of the unstable manifold of $X_\alpha$ which is initially directing to the lower-left to be the lower branch $U^{-}(0,0)$. The total unstable manifold including the two branches is denoted by: $U(0,0)=U^{+}(0,0)\bigcup U^{-}(0,0)$. Similarly, denote the part of the stable manifold (dashed curves) of $X_\beta$ which is initially directing to the upper-left to be the upper branch $S^{+}(0.5,0)$. The lower branch $S^{-}(0.5,0)$ and the total stable manifold $S(0.5,0)=S^{+}(0.5,0)\bigcup S^{-}(0.5,0)$ are denoted in the same way.

The motivation for distinguishing the upper and lower branches stems from the reflective property of $X_\alpha$. Each iteration of Eq.~\eqref{eq:two} will map $U^{+}(0,0)$ into $U^{-}(0,0)$, and vice versa. Thus points on $U(0,0)$ ``jump" between the two branches with iterations.  Define the $twice$ $iterated$ $map$ to be the compound map of two successive mappings under Eq.~\eqref{eq:two}, then $U^{+}(0,0)$ and $U^{-}(0,0)$ are invariant.  On the other hand, $X_\beta$ is non-reflective, points stay on the same branch of the stable manifold with iterations. As an example, the heteroclinic orbit segment starting from $R_{0}$ ($R_{0}\in U^{+}(0,0)\bigcap S^{+}(0.5,0)$) in Fig.~\ref{fig:one} will remain on $U^{+}(0,0)\bigcap S^{+}(0.5,0)$ for even iterations $R_{\pm2}$, $R_{\pm4}$, $\cdots$, etc.  For odd iterations it will jump to $U^{-}(0,0)\bigcap S^{+}(0.5,0)$.  In the following sections the orbit segment $\lbrace R_j: R_{j+t}\rbrace$ is considered and a method is developed to obtain large numbers of iterations in both forward and inverse directions of the mapping, which is inaccessible via straight calculations of Eq.~\eqref{eq:two} because of the exponential growth rate of initial error.  

\section{Theory and Calculations}
\label{Theory and Calculations}

\subsection{\label{ Manifold Stability}Stability in the neighborhood of invariant manifolds}

\begin{figure}[h]
\centering
{\includegraphics[width=8cm]{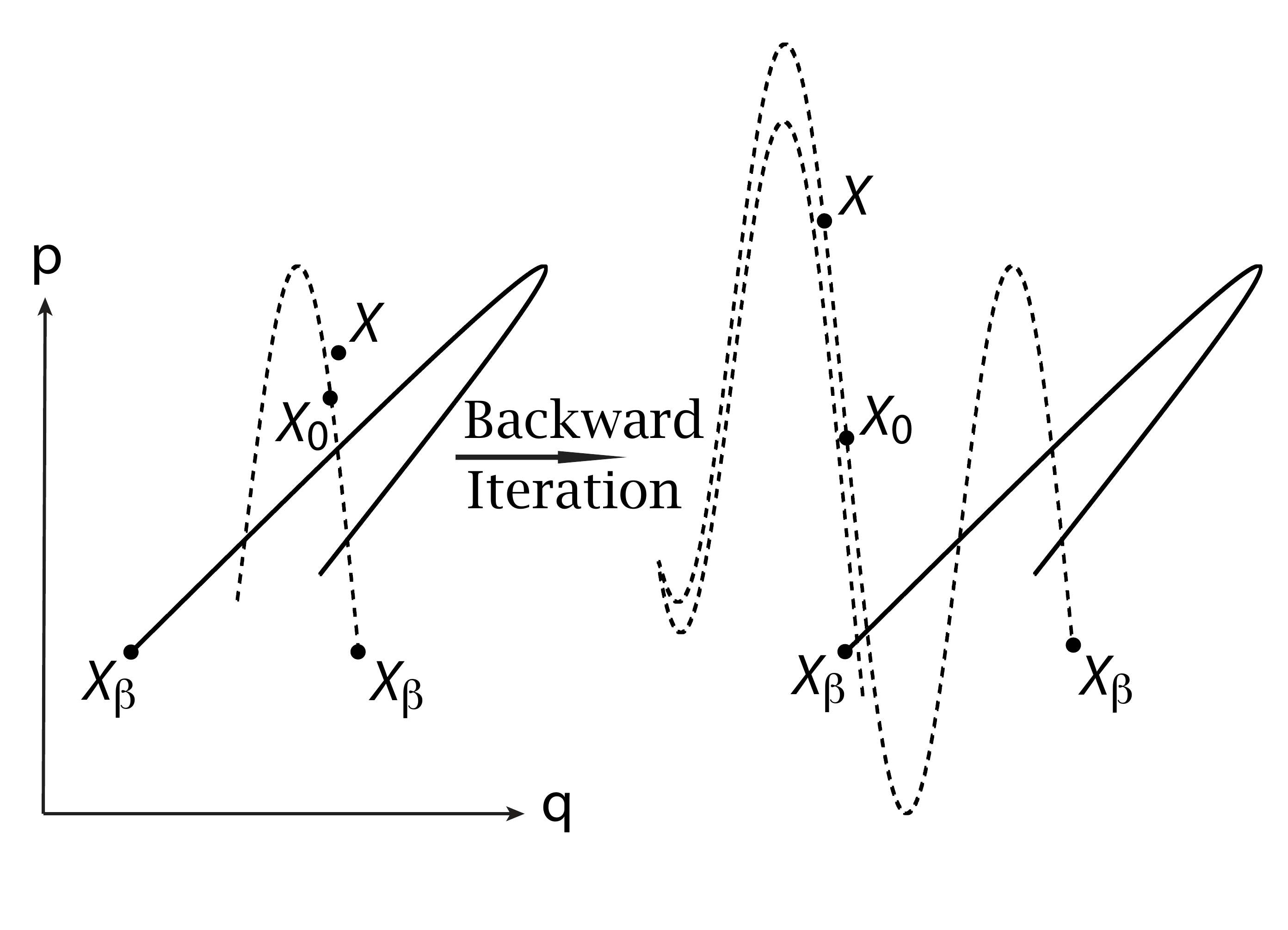}
} \caption{Manifold stability: deviation from the stable manifold collapses toward the manifold under inverse iteration, respectively.  Assume for illustration that the phase point $X_\beta$ is shown wrapped once around the torus, and is the same point in both locations.}
\label{fig:two}
\end{figure}

It is well-known that regions near stable and unstable manifolds inherit strong stability in the sense that any point in the neighborhood of the stable (unstable) manifold with a small deviation from the manifold will exponentially collapse toward the manifold under inverse (forward) iteration~\cite{Hirsch77}.  It guarantees the legitimacy of many methods to generate the global stable and unstable manifolds by propagating points from the corresponding subspace near the fixed points.  The invariant manifolds can thus be calculated to high numerical accuracy and in quite stable ways, as shown by earlier researchers facilitating different approaches \cite{You91,Hobson93,Guckenheimer93,Dellnitz96,Johnson97,Dellnitz97,Krauskopf98a,Krauskopf98b,Krauskopf99,Junge00,Krauskopf03,Mancho03,England04,Henderson05,Li12}.  A comprehensive survey of methods can be found in \cite{Krauskopf05}.  Consider the example schematically illustrated in Fig.~\ref{fig:two}, where a portion of a general homoclinic tangle commonly seen in Hamiltonian systems, formed by the unstable manifold of a non-reflective fixed point $X_\beta$ and its stable manifold is illustrated. Note that Fig.~\ref{fig:two} is not specific to the kicked rotor. Heteroclinic tangles for the kicked rotor are significantly different from Fig.~\ref{fig:two}.  However, the structural stability of the manifolds is always maintained.  Suppose the point $X_{0}$ is a true point on the manifold and $X$ is found as a result of some numerical calculation instead of $X_{0}$.  The deviation of $X$ from the stable manifold shrinks to zero under inverse iteration, making $X$ collapse exponentially toward the stable manifold. However, the tangential deviation between $X$ and $X_{0}$ along the stable manifold will be magnified exponentially, and the two points will be spaced further apart along the manifold. 

\begin{figure}[h]
\centering{\includegraphics[width=8cm]{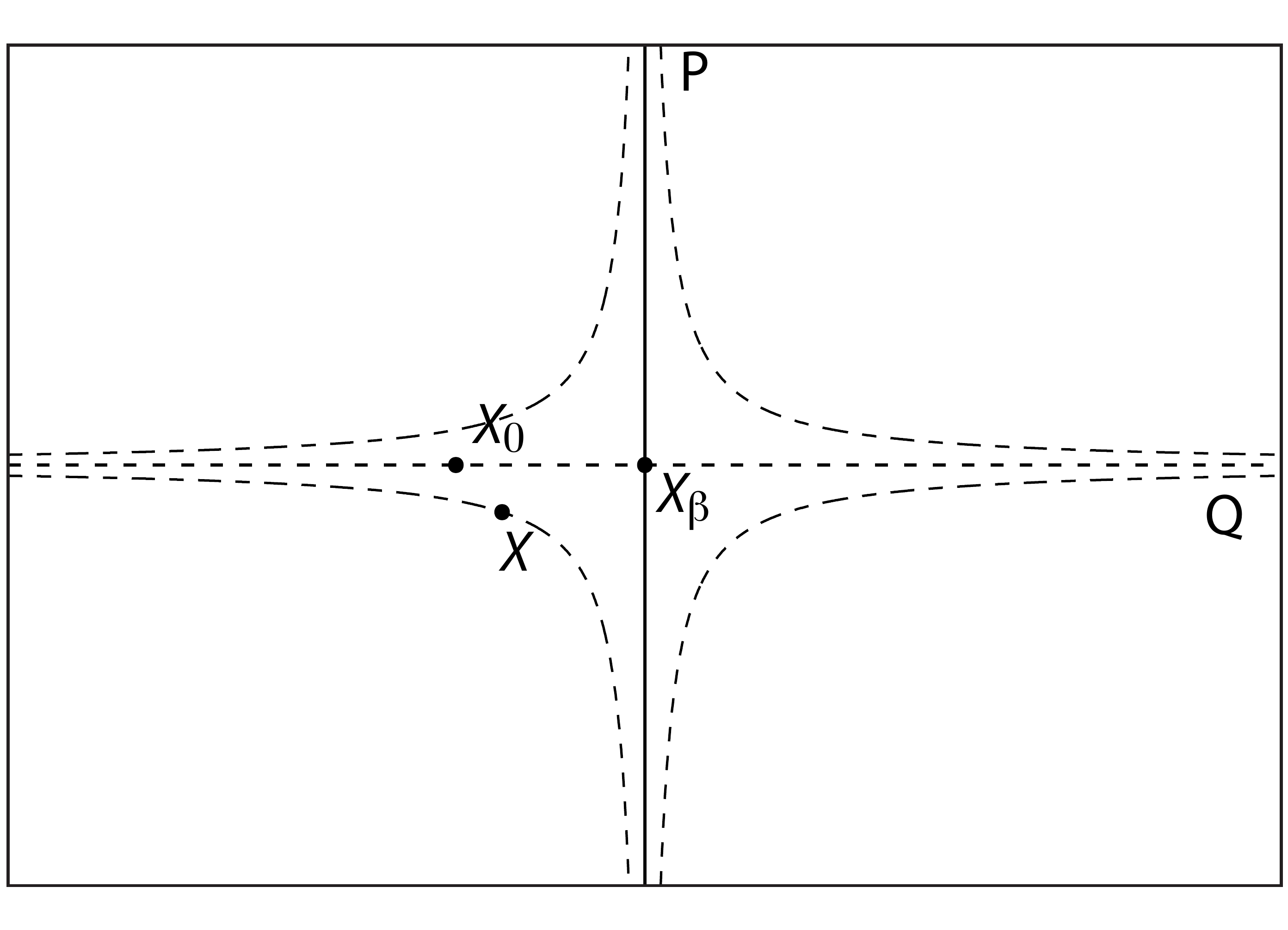}
} \caption{The image of the stable and unstable manifolds of point $X_\beta$ in the normal form coordinates. The invariant hyperbola under the mapping of Eq.~\eqref{eq:three} that passes through the point $X$ is also plotted in dashed curves.}
\label{fig:three}
\end{figure}

To show this, imagine performing a normal form transformation of the heteroclinic tangle on the left in Fig.~\ref{fig:two}. The picture in the normal form coordinates is shown in Fig.~\ref{fig:three}. The P and Q axes are the images of the unstable manifold and the stable manifold of $X_\beta$ respectively under the normal form transformation. The convergence zone of the normal form series extend along the manifolds to infinity~\cite{Silva87}.  Assuming that $X$ lies inside this convergence zone, it evolves under the equation~\cite{Silva87,Moser56},
\begin{equation}\label{eq:three}
\begin{split}
& P'=\frac{P}{U(QP)}\\
\\
& Q'=U(QP)Q
\end{split}
\end{equation}
where $(Q',P')$ is the image of $(Q,P)$ under one inverse iteration of the map. $U(QP)$ is a function that depends only on the product $QP$ and is analytic in a neighborhood of $QP=0$. Thus the product $QP$ is preserved by mapping of Eq.~\eqref{eq:three}: $Q'P'=QP$ which yields an invariant hyperbola under the iterations, traced out by the dashed curves in Fig.~\ref{fig:three}. Under successive inverse iterations, $X$ will follow the hyperbola outward to infinity, the distance between $X$ and the stable manifold will decrease by a factor of $U(QP)$ after each iteration, thus $X$ will converge to the stable manifold exponentially. Transfer back to the phase space in $(q,p)$ coordinates shows that this exponential convergence always holds~\cite{Mitchell03}.

Equivalently, it can be shown that any deviation inside the convergence zone of the unstable manifold will collapse onto the unstable manifold under forward iteration. The details are skipped here. This provides a scheme to characterize the behavior of points near the manifolds and implies the manifolds have a certain tolerance to initial errors when generated numerically.  This tolerance is the manifold stability which is the starting point of locating successive iterations of heteroclinic intersecting points. 

\subsection{Heteroclinic orbits}
\label{Heteroclinic orbit}

\begin{figure}[h]
\centering
{\includegraphics[width=8cm]{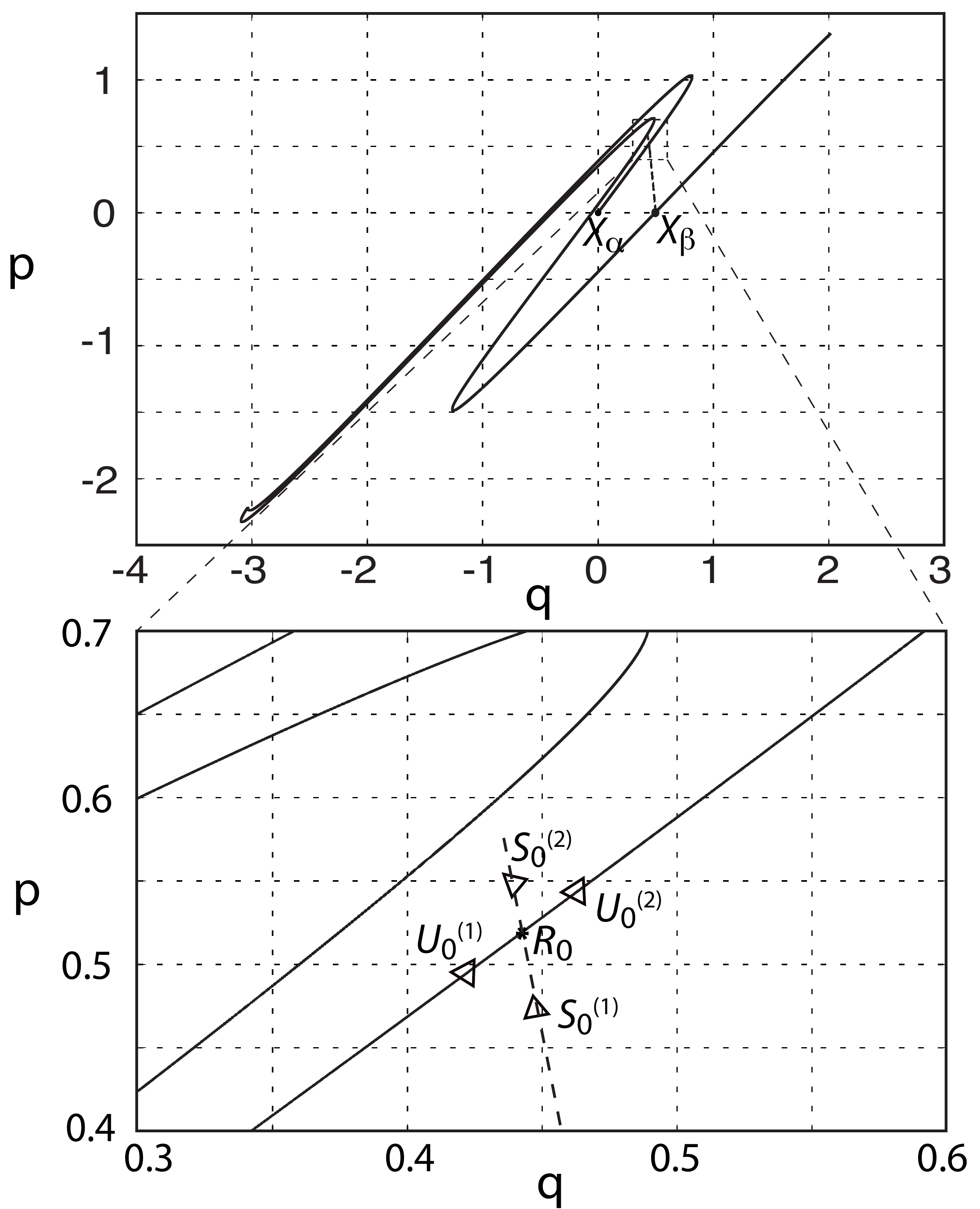}}
 \caption{Heteroclinic tangle of the kicked rotor. Shown here are $U^{+}(0,0)$ and $S^{+}(0.5,0)$. Note that they are the same manifolds shown in Fig.~\ref{fig:one}, with $U^{+}(0,0)$ extending further. The phase point $R_{0}$ in the lower zoomed-in graph is also the same point as in Fig.~\ref{fig:one}.}
\label{fig:four}
\end{figure}

The method section is illustrated by calculating the orbit segment of the heteroclinic intersecting point $R_{0}\in U^{+}(0,0)\bigcap S^{+}(0.5,0)$ shown in Fig.~\ref{fig:four}.  It facilitates a straightforward insertion technique to insert new points into the manifolds at each iteration to maintain a dense enough set of points to better interpolate the manifolds.  This kind of insertion technique (and more sophisticated ones) has been used by many authors to generate the invariant manifolds to a high accuracy on fine scales.  See for example~\cite{Hobson93} for a widely used technique which sets up a criterion examining the local curvature of the manifold and inserts points if the curvature exceeds certain thresholds.  A more elaborated method is developed in \cite{Dritschel97}, which includes a point redistribution procedure after the insertion to achieve greater resolution in certain regions of interests.  A notable generalization of these techniques into aperiodically time-dependent vector fields is introduced in \cite{Mancho03}.  Approximation to the invariant manifolds using geodesic circles and the Hobson's criterion in higher dimensional cases can be found in \cite{Krauskopf99,Krauskopf03}.  Here, only the part of the manifolds near the homoclinic intersections are needed.  Thus a linear insertion of new points at every iteration is sufficient, which is demonstrated in the following.  Define the positive (negative) direction on $U^{+}(0,0)$ to be 
pointing outward (inward) from (towards) $(0,0)$ along the manifold, and the positive (negative) direction of $S^{+}(0.5,0)$ to be pointing outward (inward) from (towards) $(0.5,0)$ along the manifold.  Calculate $R_{0}$ first by interpolation methods. Then pick two points on $U^{+}(0,0)$, namely $U_{0}^{(1)}$ and $U_{0}^{(2)}$, such that $U_{0}^{(1)}$ is adjacent to $R_{0}$ on the negative side, and $U_{0}^{(2)}$ is the point adjacent to $R_{0}$ on the positive side. Another pair of points on $S^{+}(0.5,0)$ are picked in the same way, namely $S_{0}^{(1)}$ and $S_{0}^{(2)}$, such that $S_{0}^{(1)}$ is the point adjacent to $R_{0}$ on the negative side, and $S_{0}^{(2)}$ is adjacent to $R_{0}$ on the positive side. Due to the reflective property of point $(0,0)$, the $R_{i}$ switch back and forth under each iteration as follows:\\

\noindent $R_{\pm i}\in U^{+}(0,0)\bigcap S^{+}(0.5,0)$, ($i$ even)\\
$R_{\pm i}\in U^{-}(0,0)\bigcap S^{+}(0.5,0)$, ($i$ odd).\\

\noindent The scheme is to obtain the set $\{R_{\pm i}\}$ ($i$ even) by intersecting $U^{+}(0,0)$ with $S^{+}(0.5,0)$ and iterate it once to obtain the set $\{R_{\pm i}\}$ ($i$ odd).  

Consider the inverse iterations with even iteration numbers first.  $R_{-i}$ can be calculated by iterating $R_{0}$ with the inverse map directly.  However, the maximum number of iterations is limited due to the exponential growth rate of initial error associated with $R_{0}$. Under inverse iterations, the component of the error transverse to the stable manifold will vanish by the virtue of manifold stability, but the tangential component will be magnified along the stable manifold. This will cause the calculated $R_{-i}$ to ``drift away" from the true result exponentially fast. To avoid this, instead of iterating $R_{0}$, iterate $S_{0}^{(1)}$ and $S_{0}^{(2)}$ inversely to obtain their images $S_{-i}^{(1)}$ and $S_{-i}^{(2)}$. Then locate the intersecting point between $U^{+}(0,0)$ and the straight line segment connecting $S_{-i}^{(1)}$ and $S_{-i}^{(2)}$.  They provide a better estimate of $R_{-i}$.  

\begin{figure}[h]
\centering
{\includegraphics[width=8cm]{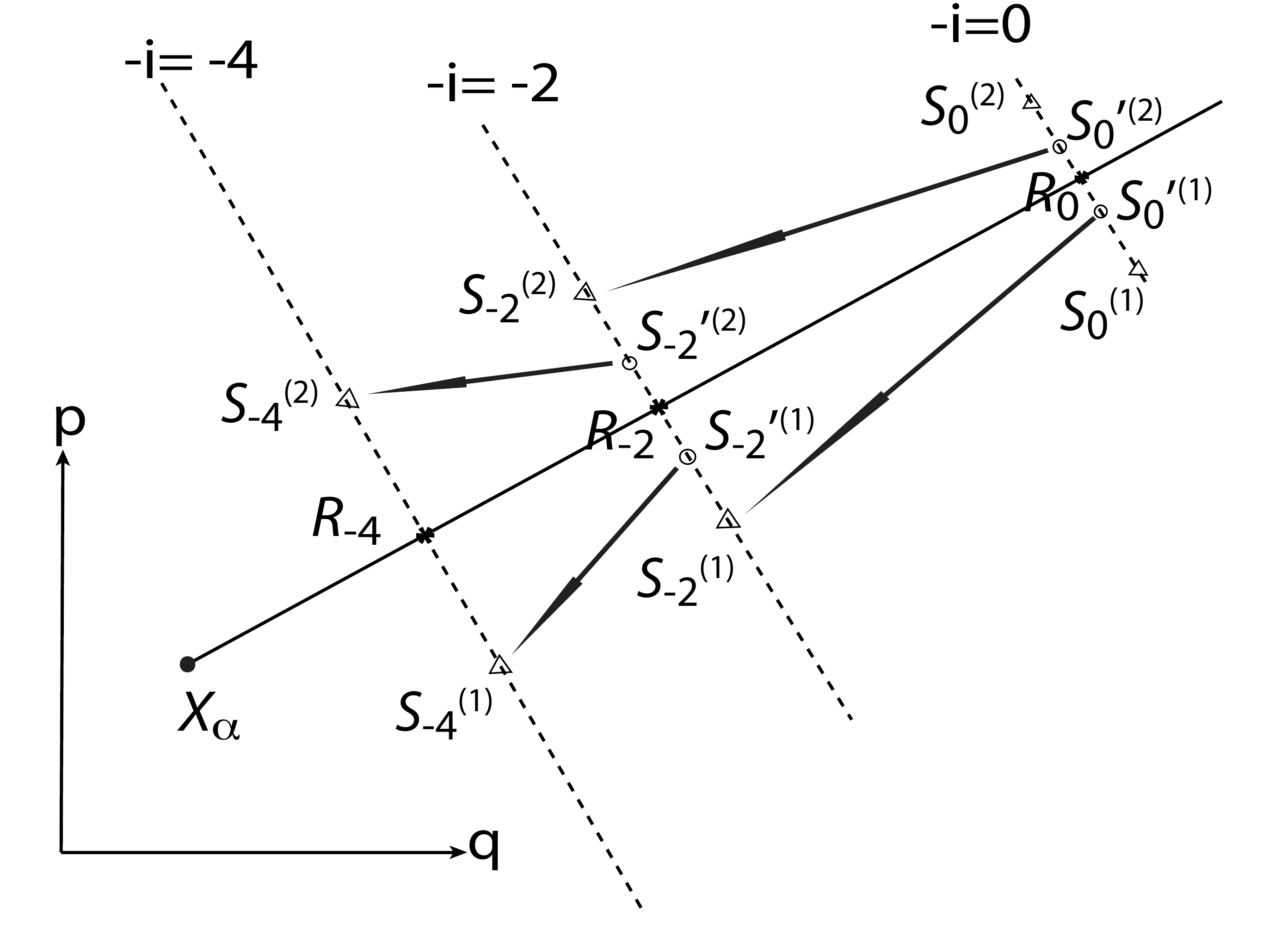}}
 \caption{Calculating $R_{-i-2}$ from $R_{-i}$.  Insert a new pair of points $S_{-i}^{\prime (1)}$ and $S_{-i}^{\prime (2)}$ near $R_{-i}$, then iterate them inversely twice to $S_{-i-2}^{(1)}$ and $S_{-i-2}^{(2)}$.  $R_{-i-2}$ is calculated as the intersection between the unstable manifold and the line segment connecting $S_{-i-2}^{(1)}$ and $S_{-i-2}^{(2)}$.  The insertion of  $S_{-i}^{\prime (1)}$ and $S_{-i}^{\prime (2)}$ is carried out in order to cancel out the exponential growth that otherwise would occur, thus ensuring a good approximation to the local stable manifold.}
\label{fig:five}
\end{figure}

There is a difficulty however, i.e.~the exponential growth of the distance between $S_{-i}^{(1)}$ and $S_{-i}^{(2)}$.  An improved procedure is illustrated in Fig.~\ref{fig:five}.
As the distance becomes larger, at some iteration the straight line connecting the two points becomes a bad approximation to the local stable manifold (even though the propagated points, $S_{-i}^{(1)}$ and $S_{-i}^{(2)}$, continue to converge towards the stable manifold).  Take the example of inverse  iterations from $R_{0}$ to $R_{-2}$. Instead of iterating $S_{0}^{(1)}$ and $S_{0}^{(2)}$ inversely to obtain $S_{-2}^{(1)}$ and $S_{-2}^{(2)}$, insert two new points $S_{0}^{\prime (1)}$ and $S_{0}^{\prime (2)}$ on the line segment connecting $S_{0}^{(1)}$ and $S_{0}^{(2)}$, such that the distance between $R_{0}$ and $S_{0}^{\prime (1)}$ is reduced by the instability $1/U^2(QP)$ relative to the distance between $R_{0}$ and $S_{0}^{(1)}$.  Do similarly with $S_{0}^{\prime (2)}$.  Then iterate $S_{0}^{\prime (1)}$ and $S_{0}^{\prime (2)}$ inversely twice to obtain a new pair of points $S_{-2}^{(1)}$ and $S_{-2}^{(2)}$.  The structural stability ensures that any error placing $S_{0}^{\prime (1)}$ and $S_{0}^{\prime (2)}$ on the stable manifold collapses exponentially.  

The orbit point $R_{-2}$ is located as the intersection between the unstable manifold and the straight line connecting $S_{-2}^{(1)}$ and $S_{-2}^{(2)}$.  Repeat the same insertion and iteration process to get $R_{-4},R_{-6},\dots$.  Practical modifications of the simple description of the process given here, such as using four or six iterations of the map instead of two or reducing the distance by more than $1/U^2(QP)$, can be introduced for particular dynamical systems and circumstances.  It is not necessary to invoke this entire procedure in order to obtain the odd iteration numbered points.  One can just iterate $R_{-i}$ once inversely to obtain $R_{-i-1}$.  The accuracy of the heteroclinic orbit is not compromised in this way.

\begin{figure}[h]
\centering
{\includegraphics[width=8.5cm]{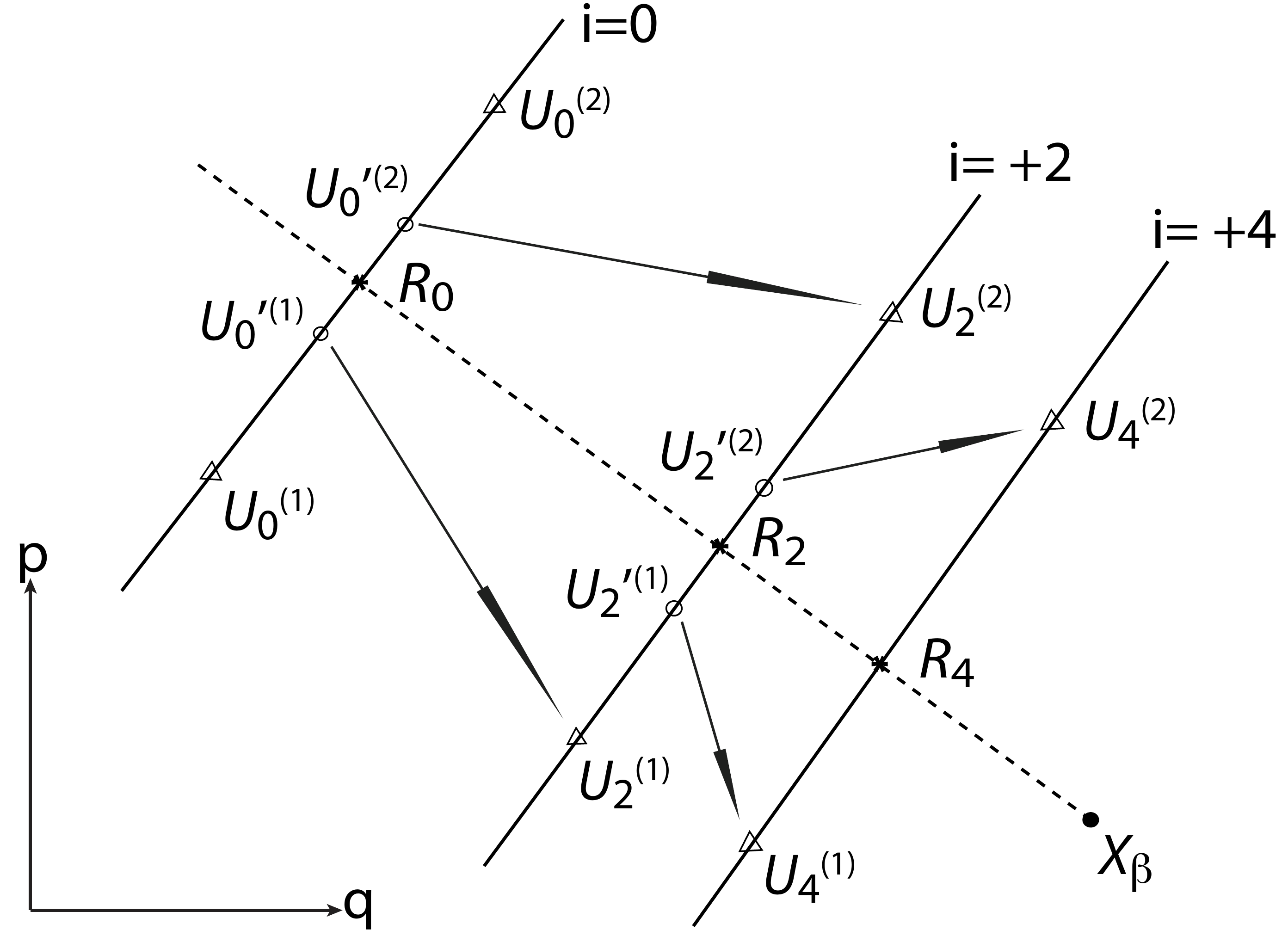}}
 \caption{Calculating $R_{i+2}$ from $R_{i}$: Insert a new pair of points $U_{i}^{\prime (1)}$ and $U_{i}^{\prime (2)}$ near $R_{i}$, then iterate them forward twice to $U_{i+2}^{(1)}$ and $U_{i+2}^{(2)}$. $R_{i+2}$ is calculated as the intersection between the stable manifold and the line segment connecting $U_{i+2}^{(1)}$ and $U_{i+2}^{(2)}$.  The insertion of  $U_{i}^{\prime (1)}$ and $U_{i}^{\prime (2)}$ cancels out the exponential deviation between $U_{i}^{(1)}$ and $U_{i}^{(2)}$, ensuring a good approximation to the local unstable manifold.}
\label{fig:six}
\end{figure}

The calculations of forward iterations are similar, the only difference is the use of unstable manifolds instead of stable ones.  The idea is to find the pair of points $U_{i}^{(1)}$ and $U_{i}^{(2)}$ by the insertion of $U_{i-2}^{\prime (1)}$ and $U_{i-2}^{\prime (2)}$ on $U^{+}(0,0)$. The scheme is shown in Fig.~\ref{fig:six}. This procedure calculates $R_{+i}$ ($i$ even), with the $R_{+i}$ ($i$ odd) calculated similarly as with the stable manifold by mapping the even iterates forward once. 

\subsection{Kicked rotor heteroclinic orbits}
\label{Numerical Results}
\begin{figure}[h]
\centering
{\includegraphics[width=8cm]{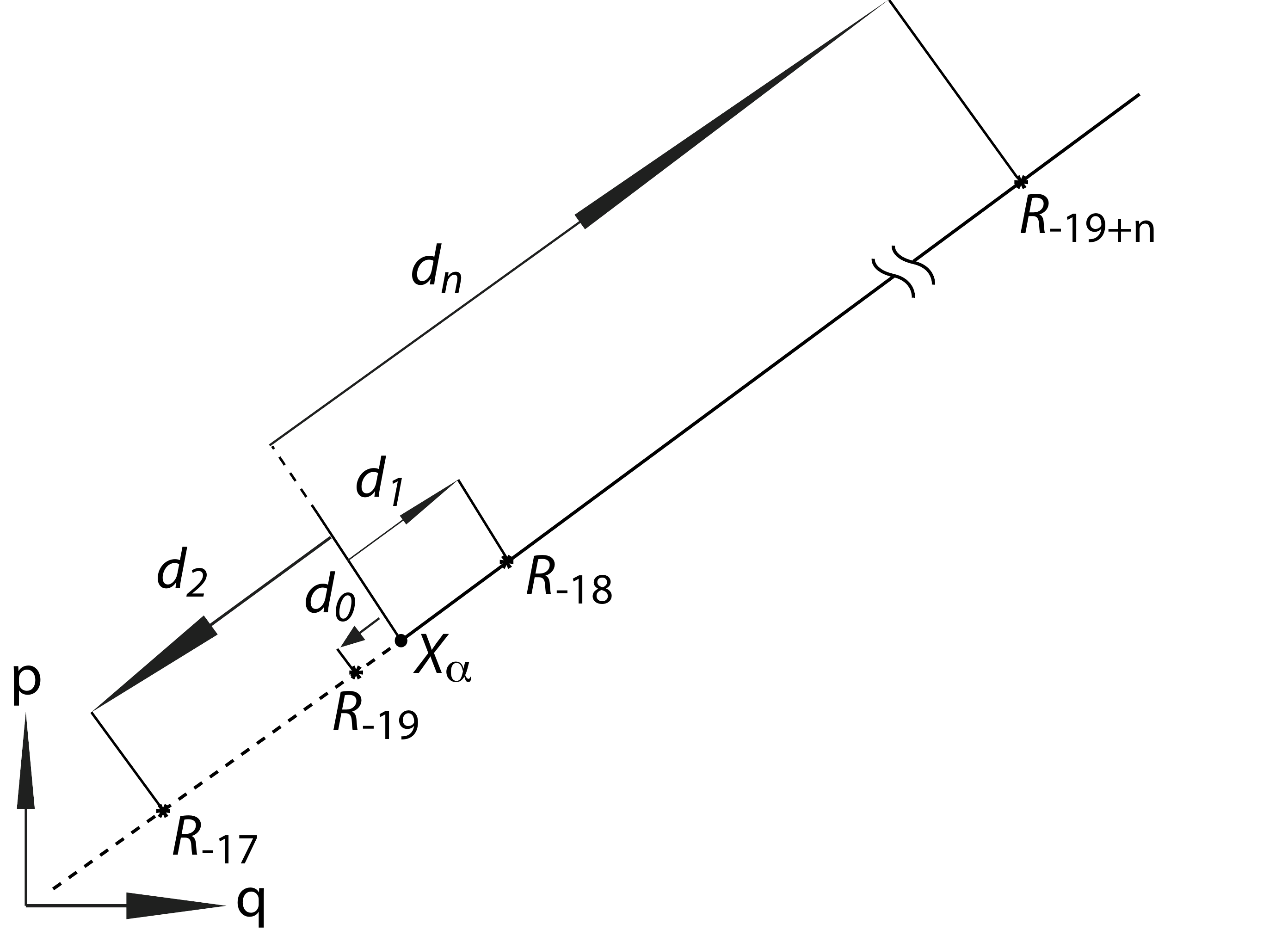}}
 \caption{Distance between forward iterations of $R_{-19}$ and $(0,0)$ (not to scale). $R_{-19}$ is marked by $d_{0}$ and $R_{-19+n}$ by $d_{n}$. Note that the integer $n$ should be an odd number since the orbit point $R_{-19+n}$ is located on the upper branch in our figure. $U^{+}(0,0)$ is the solid line while $U^{-}(0,0)$ the dashed line.}
\label{fig:seven}
\end{figure}

The above scheme applies to the heteroclinic orbit starting from $R_{0}$.  By changing the initial choice of $R_{0}$ within a single loop structure (or turnstile for the homoclinic case) and repeated use of the scheme, we can calculate a whole set of distinct heteroclinic orbits.  Orbits with longer excursion lengths only require that we generate longer segments of the stable and unstable manifolds at the beginning, which do not introduce any difficulties.  In this section numerical results are given for two example heteroclinic orbits.  The first case is given in detail for the particular orbit $R_{0}$ shown in Figs.~\ref{fig:one},\ref{fig:four}, which cannot be followed directly with the mapping forward and inversely with time in a double precision calculation better than $R_{-5}$ to $R_{5}$.  Its initial condition is well approximated as $R_{0} \approx (0.44217031018010239,0.51879785319959426)$.  It happens to be the heteroclinic orbit with the simplest excursion away from $\{X_\alpha,X_\beta\}$.  With the method just described, it is possible without any additional technical enhancements to follow this heteroclinic orbit from $R_{-19}$ to $R_{14}$.  The inverse and forward time limits are determined by the fact that the orbit expressed in double precision is indistinguishably close to $X_\alpha$ and $X_\beta$ by those times.  As the orbit approaches fixed limiting points in its past and future exponentially quickly, it is not possible to illustrate it very well in a simple phase space plot with successive points labeled $\{R_{-19}\cdots R_{14}\}$.  Instead, a schematic is shown of the first few iterations beginning with $R_{-19}$ in Fig.~\ref{fig:seven}.

To show that the solution is behaving properly, consider the stability analysis of the initial and final fixed points.  In a small neighborhood of either $X_\alpha=(0,0)$ or $X_\beta=(0.5,0)$, the mapping can be approximated by the linearized tangential equation $\delta X_{\alpha (\beta)}(n+1)=M_{X_{\alpha (\beta)}}\delta X_{\alpha (\beta)}(n)$, where $\delta X_{\alpha (\beta)}(n)=(\delta q_{n},\delta p_{n})_{\alpha (\beta)}^{T}$ is the deviation of the $n^{th}$ iteration relative to $X_{\alpha (\beta)}$, and $M_{X_{\alpha (\beta)}}$ is the stability (Jacobi) matrix of $X_{\alpha (\beta)}$.  A straightforward calculation of the stability matrices for both $X_{\alpha (\beta)}$ yields:
\begin{eqnarray}
\delta X_\alpha (n+1) &=& \begin{pmatrix} 1-K & 1\\-K & 1 \end{pmatrix} \delta X_\alpha (n) \nonumber \\
\delta X_\beta (n+1) &=&\begin{pmatrix} 1+K & 1\\ K & 1 \end{pmatrix} \delta X_\beta (n)\ .
\end{eqnarray}
The eigenvalues associated with the stable manifolds of $X_{\alpha (\beta)}$ are $\lambda^{s}_{\alpha (\beta)}=\left(2\pm K \mp\sqrt{K^{2} \pm 4K}\right)/2$, where the lower sign refers to $\alpha$ and the upper sign to $\beta$.  Likewise, the eigenvalues associated with the unstable manifolds are $\lambda^{u}_{\alpha (\beta)}=(2\pm K \pm\sqrt{K^{2}\pm 4K})/2$.  Setting $K=8.25$ gives the four values relevant to this orbit example.

\begin{figure}[h]
\centering
{\includegraphics[width=8cm]{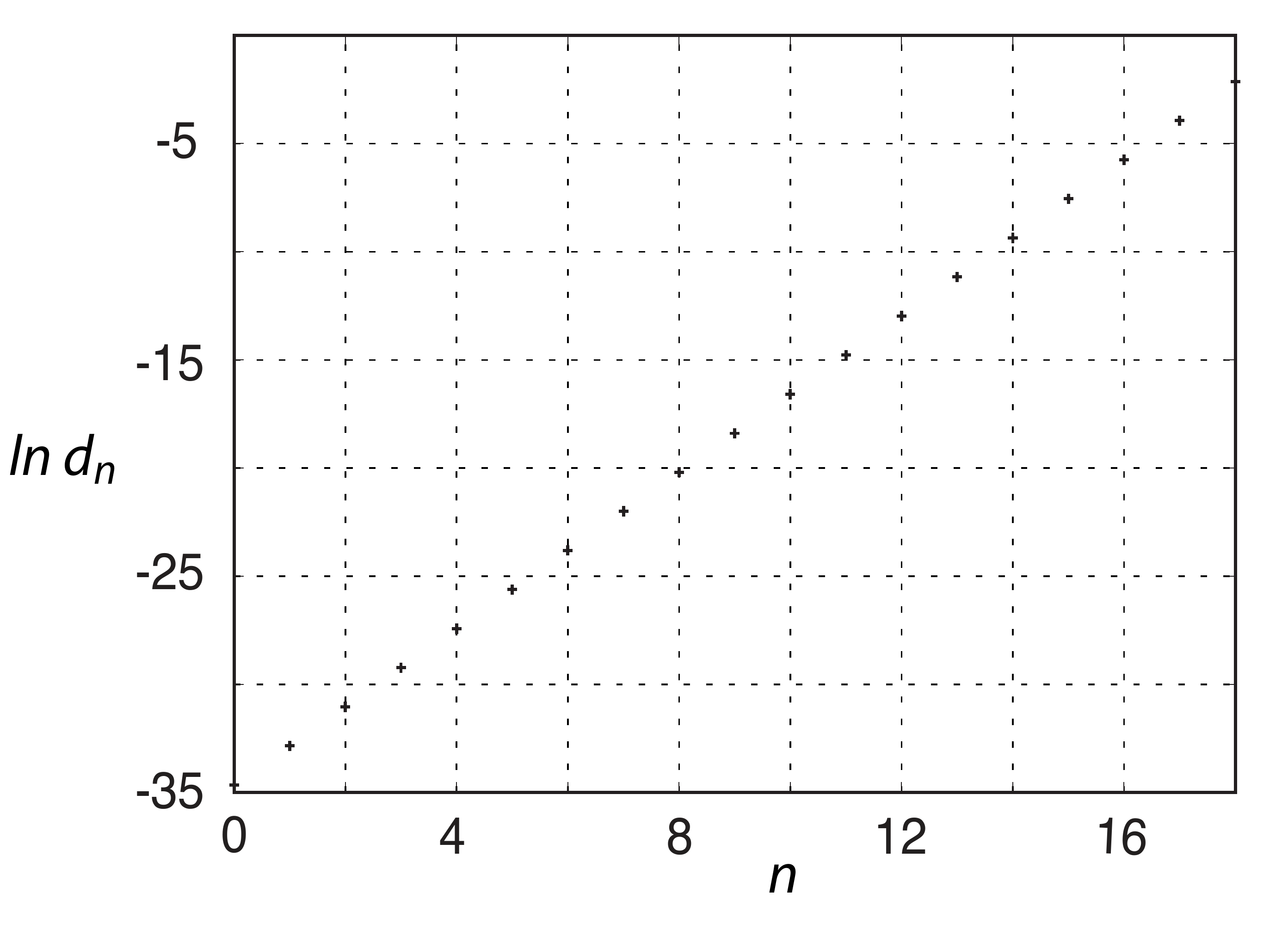}}
 \caption{Exponential convergence under inverse iteration near $X_\alpha$. The slope of the graph matches the characteristic exponent to five decimal places; $\ln|\lambda^{u}_\alpha|=1.80593$ versus $1.80592$ for the trajectory.}
\label{fig:eight}
\end{figure}

For the early iterations $R_{n}$, $n$ close to $-19$, consider the norm of the difference between $R_{n}$ and $X_\alpha$, $d_{n}=\sqrt{\delta q_{n}^{2}+\delta p_{n}^{2}}$.  It should turn out that 
\begin{equation}
d_{n+1}\approx |\lambda^{u}_\alpha |d_{n}
\end{equation}
is an excellent approximation to the true dynamics of the heteroclinic orbit.  Following the orbit for multiple iterations as in Fig.~\ref{fig:seven} and considering the natural logarithm of the norm leads to
\begin{equation}
\ln d_{n}\approx n\ln|\lambda^{u}_\alpha |+\ln d_{0}
\end{equation}
Therefore a plot of $\ln d_{n}$ versus $n$ should be nearly a straight line with slope $\ln|\lambda^{u}_\alpha|$. Shown in Fig.~\ref{fig:eight} is the graph plotted from our calculations of the orbit segment using the manifold intersections method.  The points do indeed line up with the correct slope to a high degree of precision.  

Similar verification can be done for the forward iterations approaching $X_\beta$.  Start with the point $R_{14}$, which is the furthest forward iteration found, and count backward.  Let $d_{0}$ be the distance between $R_{14}$ and $X_\beta$, $d_{n}$ the distance between $R_{14-n}$ and $X_\beta$.  The same relation applies as before except using $|\lambda_\beta^u|$.  The plot looks identical to that of Fig.~\ref{fig:eight}, so is not included here.  The only difference is that it has a slope of $2.31808$, which is to be compared with $\ln |\lambda_\beta^u |=2.31762$.  This also differs in the fifth significant digit.

As mentioned in the introduction, in semiclassical theory classical action differences divided by $\hbar$ control quantum interferences.  Thus, even very small errors in the calculation of action differences, depending on the value of $\hbar$, can ruin the quality of a semiclassical approximation.  Exploiting the exact relationship relating the action difference between two heteroclinic orbits to the geometric area under the invariant manifolds that connect them~\cite{MacKay84a, Meiss92} gives a meaningful measure of the quality of an orbit's calculation.  

\begin{figure}[h]
\centering
{\includegraphics[width=8cm]{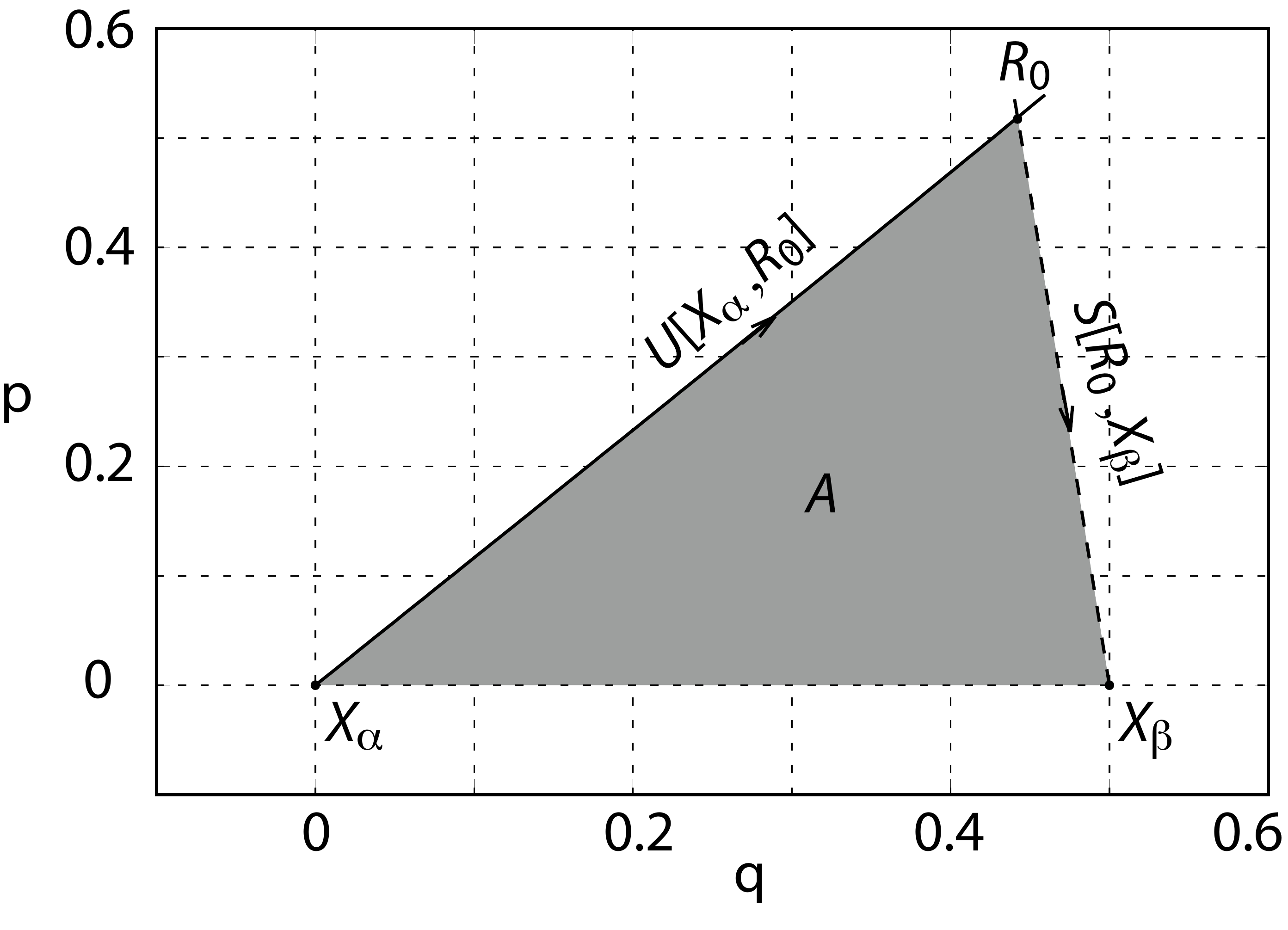}}
 \caption{Path of integration to calculate the geometric areas. $U[X_{\alpha},R_{0}]$ is the segment of the unstable manifold from $X_\alpha$ to $R(0)$. $S[R_{0},X_{\beta}]$ is the segment of the stable manifold from $R_{0}$ to $X_\beta$.  The shaded region marked `A' shows the corresponding area.}
\label{fig:nine}
\end{figure}

For the kicked rotor, the action function can be written explicitly as:
\begin{equation}
F(X_{n},X_{n+1})=\frac{(q_{n+1}-q_{n})^{2}}{2}+\frac{K}{4\pi^{2}}\cos2\pi q_{n}
\end{equation}
where $X_{n}=(q_{n},p_{n})$ is an arbitrary phase point that is mapped to some $X_{n+1}=(q_{n+1},p_{n+1})$ under Eq.~(\ref{eq:two}) without the $mod$ operation.  If used as an $F_1(q,Q;t)$ generating function, $-F$ generates such a map.  The two classical action functions $F(X_{\alpha (\beta)},X_{\alpha (\beta)})=\pm K/4\pi^2$ are invariant under translation in time as they refer to fixed points of the map.  Repeated use of Eq.~(5.6) from~\cite{Meiss92} gives:
\begin{equation}
\sum_{i=-\infty}^{0}\Bigg[F(R_{i-1},R_{i})-F(X_{\alpha},X_{\alpha})\Bigg]=\int\limits_{U[X_{\alpha},R_{0}]}p\mathrm{d}q
\end{equation} 
The path of integration $U[X_{\alpha},R_{0}]$ is the segment of unstable manifold from $X_\alpha$ to $R_{0}$, as shown in Fig.~\ref{fig:nine}. The direction is denoted by an arrow.
Similarly:
\begin{equation}
\sum_{i=0}^{+\infty}\Bigg[F(R_{i},R_{i+1})-F(X_{\beta},X_{\beta})\Bigg]=\int\limits_{S[R_{0},X_{\beta}]}p\mathrm{d}q
\end{equation}
The path of integration $S[R_{0},X_{\beta}]$ is the segment of stable manifold from $R_{0}$ to $X_\beta$, as shown in Fig.~\ref{fig:nine}.
Combining the above two equations gives:
\begin{equation}\label{eq:thirteen}
\begin{split}
\sum_{i=-\infty}^{0}\Bigg[F(R_{i-1},R_{i})-F(X_{\alpha},X_{\alpha})\Bigg]+\\
\sum_{i=0}^{+\infty}\Bigg[F(R_{i},R_{i+1})-F(X_{\beta},X_{\beta})\Bigg]\\
=\int\limits_{U[X_{\alpha},R_{0}]}p\mathrm{d}q+\int\limits_{S[R_{0},X_{\beta}]}p\mathrm{d}q
\end{split}
\end{equation}
For the classical action calculation, the infinite sums are truncated to $R_{-19}$ and $R_{14}$, which are as close to the fixed points as a straightforward double precision calculation can be.  The classical action difference gives $0.12938887802084850$, whereas a construction of the unstable and stable manifold segments and numerical integration of the area denoted `A' in Fig.~\ref{fig:nine} gives $0.12938887802085794$.  They differ in the $14^{th}$ decimal place, the limit of double precision calculations.

\begin{figure}[h]
\centering
{\includegraphics[width=8cm]{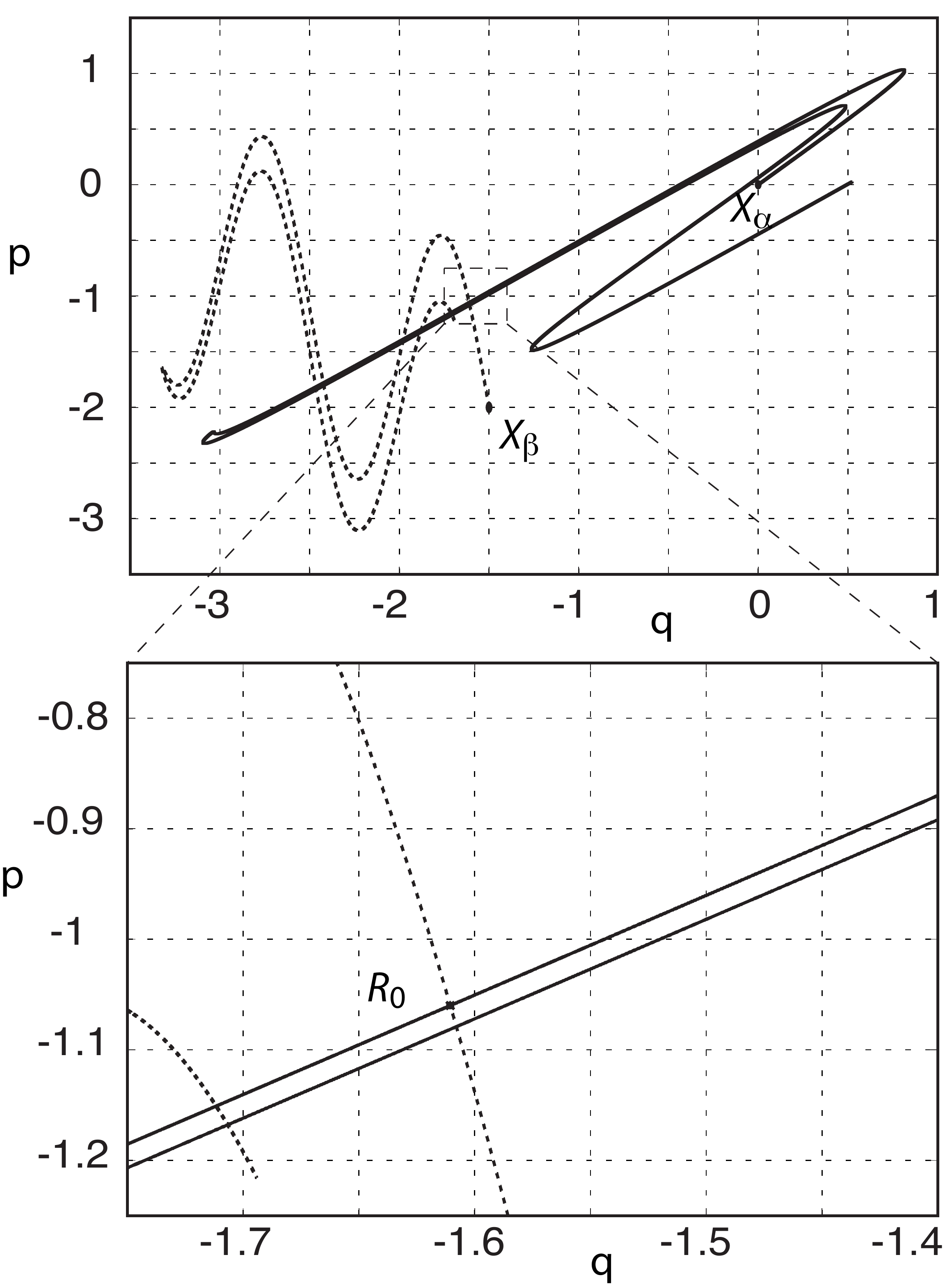}}
 \caption{Heteroclinic intersection $R_{0}$ between $U^{+}(0,0)$ and $S^{+}(-1.5,-2)$.  Note that $S^{+}(-1.5,-2)$ is just $S^{+}(0.5,0)$ (same $X_\beta)$ shifted by the winding numbers $(-2,-2)$. Its initial values are $R_{0}= (-1.6105430949740283,-1.0599500106337416)$.}
\label{fig:ten}
\end{figure}

A more complicated heteroclinic orbit is denoted with the point $R_{0}$ shown in Fig.~\ref{fig:ten}.
It not only makes a more complicated excursion away from $\{X_\alpha,X_\beta\}$, it also wraps around both directions of the unfolded torus, and thus translates from one phase space copy to another.  The peculiarity of this kind of heteroclinic orbit is that its fixed point $X_\beta$ with nonzero $n_{p}$-winding number is constantly shifting in the $q$ coordinate on the phase plane tiled with unfolded tori under iteration, along with its stable and unstable manifolds when the map is applied.  This follows from the mapping equations, Eq.~(\ref{eq:two}), by dropping the $mod$ operation.
In this case, the point  $(-1.5,-2)$ is mapped to $(-1.5-2n,-2)$ under $n$ iterations.  The stable and unstable manifolds of it are shifted the same way. Thus: \\

$R_{0}\in S^{+}(-1.5,-2)\bigcap U(0,0)$,\\

$R_{n}\in S^{+}(-1.5-2n,-2)\bigcap U(0,0)$\\

\noindent It was possible to obtain inverse iterations up to $R_{-17}$ and forward iterations up to $R_{13}$.  Similar to the previous case, the values of $\ln d_{n}$ versus $n$ compared quite closely with the characteristic exponents of $X_\alpha$ and $X_\beta$.  The beginning slope from the values of $\ln d_{n}$ is $1.80589$, which is the same to 5 decimal places with $\ln |\lambda^{u}_{\alpha}|=1.80594$.  Likewise, but a bit less accurately, the later slope from the values of $\ln d_{n}$ is $2.32168$, which is roughly the same to 4 decimal places with $\ln |\lambda^{u}_{\beta}|=2.31762$.

In order to study the area-action relation for this orbit, a slight modification of the algorithm is needed to account properly for the shifting of the phase space unfolded torus.  Let the path $U[X_{\alpha},R_{0}]$ be the segment of $U^{+}(0,0)$ from $X_\alpha$ to $R(0)$, and the path $S[R_{0},X_{\beta}]$ be the segment of $S^{+}(-1.5,-2)$ from $R_{0}$ to $X_\beta = (-1.5,-2)$.  For the $U[X_{\alpha},R_{0}]$ path of the heteroclinic orbit, there is no definitional change from the previous case, just the use of a different orbit; i.e.~$F(R_{i-1},R_{i})$ is the action function evaluated from $R_{i-1}$ to $R_{i}$ and $F(X_{\alpha},X_{\alpha})$ is the action function of the fixed point $X_{\alpha}$, independent of $i$ and equal to $K/4\pi^2$.

For the latter half of the path, i.e.~$S[R_{0},X_{\beta}]$, let $F(X_{\beta},X_{\beta})$ be the action function that maps $X_\beta=(-1.5-2i,-2)$ to $X_\beta=(-1.5-2(i+1),-2)$ as the  map takes the point $R_{i}$ to $R_{i+1}$.  The modified action is
 \begin{equation}
F(X_{\beta},X_{\beta})=2-\frac{K}{4\pi ^{2}},\forall i.
\end{equation}
The sum of the actions that contribute to double precision are:
\begin{multline}
\sum_{i=-16}^{0}\Bigg[F(R_{i-1},R_{i})-F(X_{\alpha},X_{\alpha})\Bigg]+\\
\sum_{i=0}^{+12}\Bigg[F(R_{i},R_{i+1})-F(X_{\beta},X_{\beta})\Bigg]\\
=0.16465128640816951\nonumber
\end{multline}

\begin{figure}[h]
\centering
{\includegraphics[width=8cm]{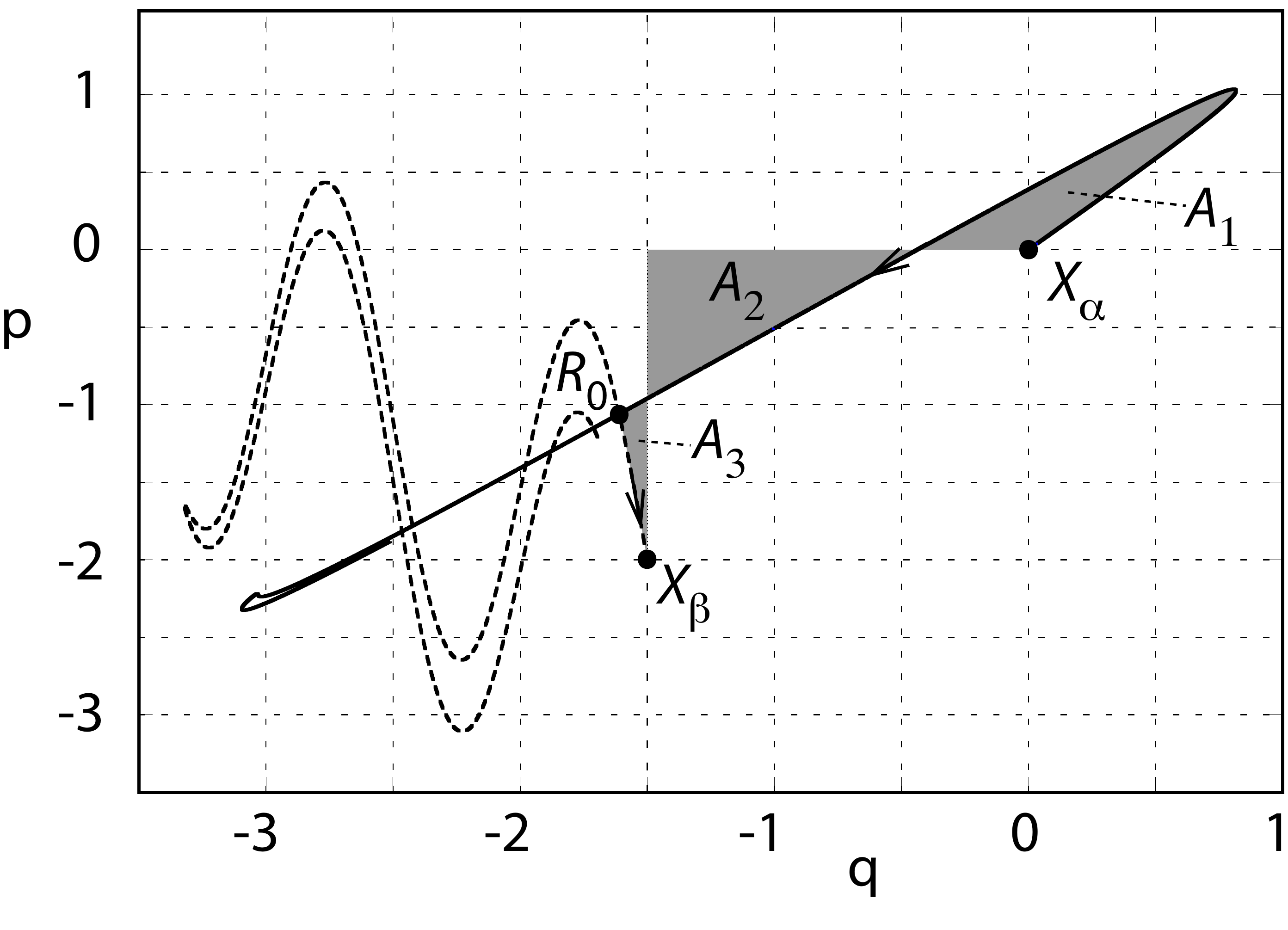}}
 \caption{Relevant areas $A_{1}$, $A_{2}$ and $A_{3}$ for the action calculation. $U[X_{\alpha},R_{0}]$ is the segment of $U^{+}(0,0)$ from $X_{\alpha}$ to $R_{0}$. $S[R_{0},X_{\beta}]$ is the segment of $S^{+}(-1.5,-2)$ from $R_{0}$ to $X_{\beta}$. The path integration $\int\limits_{U[X_{\alpha},R_{0}]+S[R_{0},X_{\beta}]}p\mathrm{d}q$ is the algebraic area $-A_{1}+A_{2}-A_{3}$.  }
\label{fig:eleven}
\end{figure}
The geometric area for this orbit is less straightforward to conceptualize.  It is shown in Fig.~\ref{fig:eleven} for clarity. Construction of the stable and unstable manifolds and using a numerical integration of the relevant areas of $-A_{1}+A_{2}-A_{3}$ gives a total of:
\begin{equation}
\begin{split}
\int\limits_{U[X_{\alpha},R_{0}]+S[R_{0},X_{\beta}]}p\mathrm{d}q=&-A_{1}+A_{2}-A_{3}\\
&=0.16465128641878213\nonumber
\end{split}
\end{equation}
Deviation of the two numbers begins in the $11^{th}$ decimal place.  This is not quite as accurate as the first heteroclinic orbit case, the simpler one.  However, the area integral is more difficult to calculate accurately and it is not known which source of possible error, either from the orbit action sum or the numerical area calculation, gives rise to the increased inaccuracy.  Nevertheless, the accuracy is excellent.

\subsection{Action relations between fixed points}
\label{actionrelations}

Here a general formula relating the action difference between a pair of fixed points to a region bounded by certain segments of the unstable and stable manifolds is derived.  The result obtained here is independent of the context of the kicked rotor and applies to any Hamiltonian system with a heteroclinic tangle.  
    
Equation \eqref{eq:thirteen} corresponds to the case that the switching point from the unstable to the stable manifold along the integration path is chosen to be $R_{0}$.  In general, it is possible to change the switching point to other orbit points $R_{k}$ from $\lbrace R_{0}\rbrace$, and modify the integration paths correspondingly.  Thus a more general relationship can be given:
 \begin{equation}\label{eq:fourteen}
\begin{split}
\sum_{i=-\infty}^{k}\Bigg[F(R_{i-1},R_{i})-F(X_{\alpha},X_{\alpha})\Bigg]+\\
\sum_{i=k}^{+\infty}\Bigg[F(R_{i},R_{i+1})-F(X_{\beta},X_{\beta})\Bigg]\\
=\int\limits_{U[X_{\alpha},R_{k}]}p\mathrm{d}q+\int\limits_{S[R_{k},X_{\beta}]}p\mathrm{d}q
\end{split}
\end{equation}

Calculating the difference between Eq.~\eqref{eq:thirteen} and Eq.~\eqref{eq:fourteen} gives the action difference between the two fixed points $X_{\alpha}$ and $X_{\beta}$.  Subtracting Eq.~\eqref{eq:thirteen} from Eq.~\eqref{eq:fourteen} leads to:
\begin{equation}\label{eq:fifteen}
\begin{split}
k\cdot\Bigg[F(X_{\beta},X_{\beta})-&F(X_{\alpha},X_{\alpha})\Bigg]\\
=\int\limits_{U[X_{\alpha},R_{k}]}p\mathrm{d}q&-\int\limits_{U[X_{\alpha},R_{0}]}p\mathrm{d}q\\
&+\int\limits_{S[R_{k},X_{\beta}]}p\mathrm{d}q-\int\limits_{S[R_{0},X_{\beta}]}p\mathrm{d}q
\end{split}
\end{equation}
For the cases that both $X_{\alpha}$ and $X_{\beta}$ are non-reflective, $U[X_{\alpha},R_{k}]$ and $U[X_{\alpha},R_{0}]$ are segments belonging to the same branch of the unstable manifold of $X_{\alpha}$, therefore:
\begin{equation}\label{eq:sixteen}
\int\limits_{U[X_{\alpha},R_{k}]}p\mathrm{d}q-\int\limits_{U[X_{\alpha},R_{0}]}p\mathrm{d}q=\int\limits_{U[R_{0},R_{k}]}p\mathrm{d}q
\end{equation}
and similarly:
\begin{equation}\label{eq:seventeen}
\int\limits_{S[R_{k},X_{\beta}]}p\mathrm{d}q-\int\limits_{S[R_{0},X_{\beta}]}p\mathrm{d}q=\int\limits_{S[R_{k},R_{0}]}p\mathrm{d}q
\end{equation}
Equation \eqref{eq:fifteen} simplifies to:
\begin{equation}\label{eq:eighteen}
\begin{split}
k\cdot\Bigg[F(X_{\beta},&X_{\beta})-F(X_{\alpha},X_{\alpha})\Bigg] \\
&=\int\limits_{U[R_{0},R_{k}]}p\mathrm{d}q+\int\limits_{S[R_{k},R_{0}]}p\mathrm{d}q
\end{split}
\end{equation}
 which holds true for Hamiltonian systems with non-reflective fixed points.  
 
\begin{figure}[h]
\centering
{\includegraphics[width=8cm]{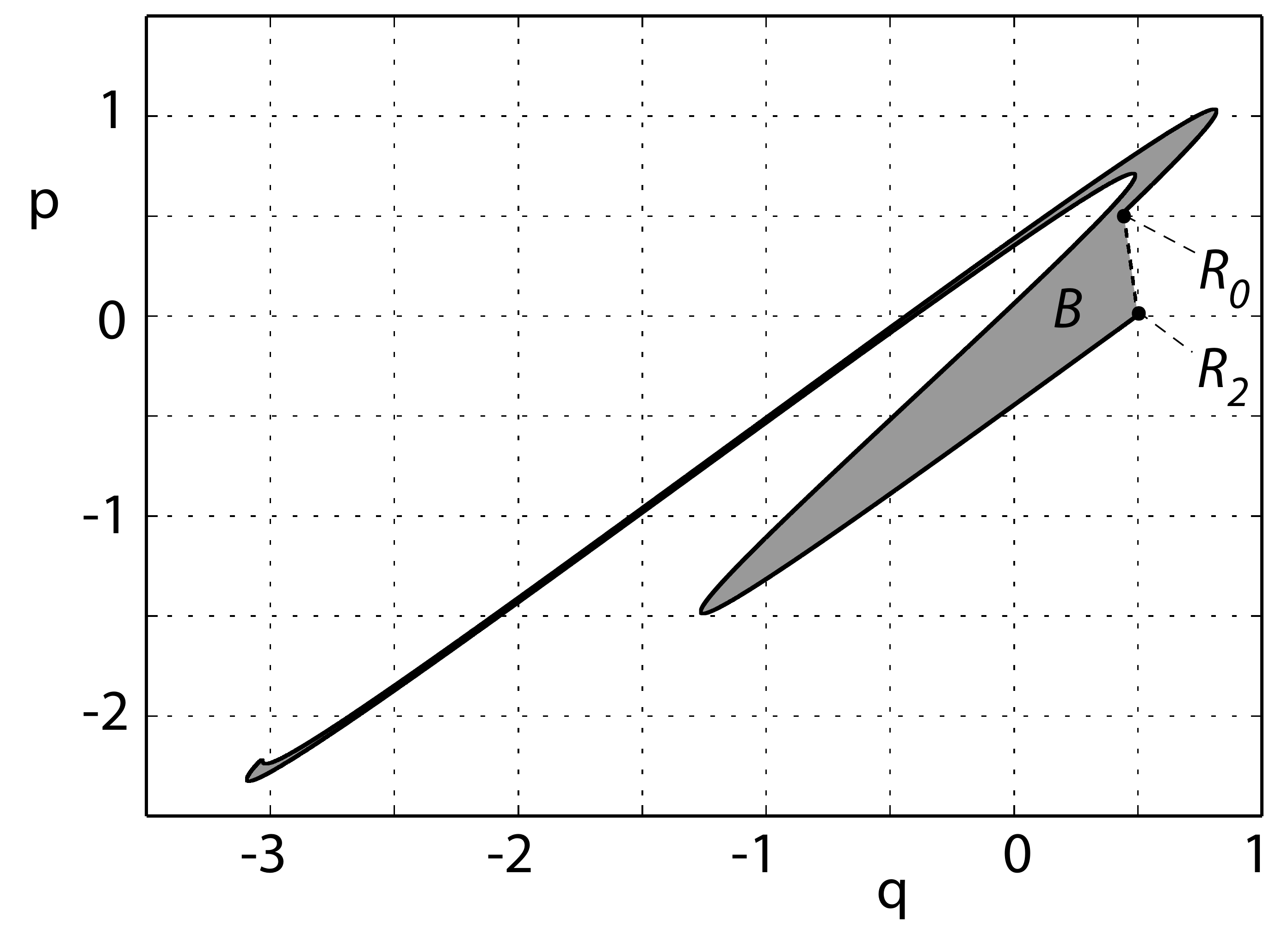}}
 \caption{The phase space region enclosed by $U[R_{0},R_{2}]$ and $S[R_{2},R_{0}]$ is labeled by $B$.  Since $U[R_{0},R_{2}]$ and $S[R_{2},R_{0}]$ are the fundamental segments of the twice iterated map, in this heteroclinic tangle, $B$ plays the role of the turnstile structure commonly seen in homoclinic tangles. Note that $R_{2}$ is extremely close to $X_{\beta}$, making it hard to distinguish between the two in the scale of the figure.}
\label{fig:twelve}
\end{figure}

In the presence of reflective fixed points, such as the kicked rotor, $U[X_{\alpha},R_{k}]$ and $U[X_{\alpha},R_{0}]$ are on the same branch only when $k$ is an even number (twice iterated map).  Letting $k=2$, we have:
\begin{equation}\label{eq:nineteen}
\begin{split}
\int\limits_{U[R_{0},R_{2}]}p\mathrm{d}q+&\int\limits_{S[R_{2},R_{0}]}p\mathrm{d}q=-B\\
=2\cdot\Bigg[&F(X_{\beta},X_{\beta})-F(X_{\alpha},X_{\alpha})\Bigg]\\
&\qquad{}=-\frac{K}{\pi^{2}}
\end{split}
\end{equation}
where the $B$ is the area of the phase space region enclosed by $U[R_{0},R_{2}]$ and $S[R_{2},R_{0}]$ shown in Fig.~\ref{fig:twelve}.  Notice that the union of $U[R_{0},R_{2}]$ and $S[R_{2},R_{0}]$ gives the $fundamental$ $loop$ $structure$ of the heteroclinic tangle under the twice iterated map.  This is in the sense that the loop is the smallest  
object which can be used to be mapped forward and inversely in order to generate the full heteroclinic tangle, $U^{+}(0,0)$ and $S^{+}(0.5,0)$, and on the loop each distinct heteroclinic orbit has only one intersection point.  In the common case of homoclinic tangles, the fundamental loop form a turnstile structure which was extensively used to study the flux entering and leaving certain phase space regions~\cite{MacKay84a,Rom-Kedar90,Wiggins92}.  However, in our case the turnstile structure is replaced by the loop shown in Fig.~\ref{fig:twelve}, which is topologically equivalent to a circle.   The area of the loop is equal to twice the action difference between the fixed points.  Thus, the loop structure of the heteroclinic tangle gives the action difference (or multiple thereof) of two periodic orbits more generally.  Numerical calculation of the area gives:
\begin{equation}
-B=-0.83589976498504137
\end{equation}
and 
\begin{equation}
-\frac{K}{\pi^{2}}=-0.83589976504928665
\end{equation}
They match up to the $11^{th}$ decimal place, which verifies Eq.~\eqref{eq:nineteen}.

\subsection{Heteroclinic fundamental loop structure}
\label{loop structure}

There is necessarily a marked difference between the heteroclinic fundamental loop structure and the homoclinic fundamental loop structure (the turnstile). \begin{figure}[h]
\centering
{\includegraphics[width=8cm]{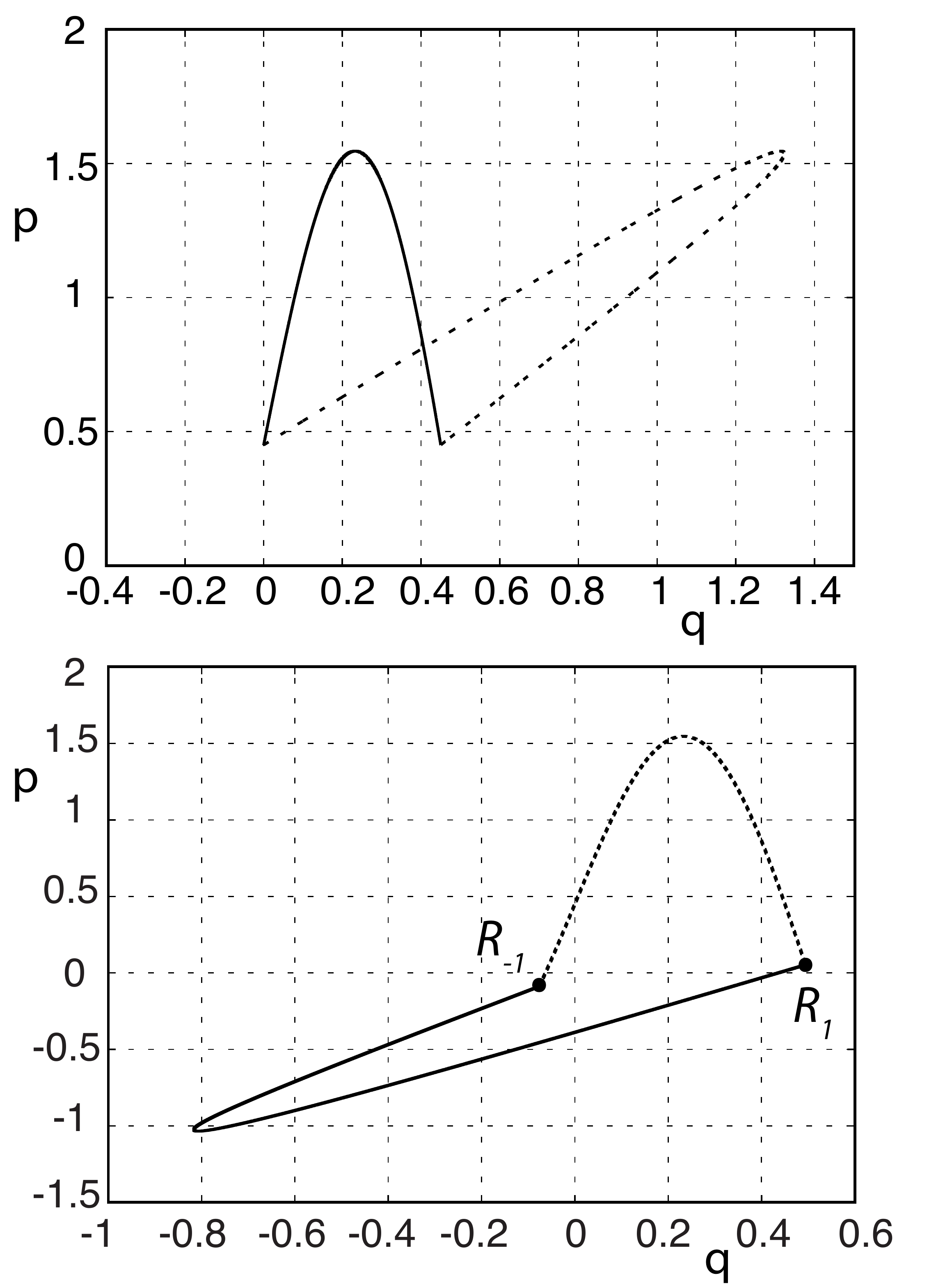}}
 \caption{The top panel shows the turnstile of the homoclinic tangle formed by $U^{+}(-0.5,0)$, $S^{+}(0.5,0)$ ,$S^{-}(-0.5,0)$ and $U^{-}(0.5,0)$, whose two lobe areas must cancel.  Structures like this are closely linked to the transport between different regions of phase space and are frequently mentioned in research literature.  The bottom panel, which shows the fundamental loop of the heteroclinic tangle between $(0,0)$ and $(0.5,0)$, is quite different.  It doesn't form a turnstile shape and is equivalent to a single circle.}
\label{fig:thirteen}
\end{figure}  It is well-known that the fundamental loop of the homoclinic tangle must have at least one crossing point between the stable and unstable manifolds in order to cancel out the flux in and out of a certain phase region~\cite{MacKay84a,Rom-Kedar90,Wiggins92}.  Using the action relations between the fixed points of Subsection~\ref{actionrelations} implies the same conclusion.  Furthermore, for the case of the heteroclinic tangle, the fundamental loop doesn't even need a crossing point between the stable and unstable manifolds, as shown below. 

For both the case without reflection (Eq.~\eqref{eq:eighteen}) and the case with reflection (Eq.~\eqref{eq:nineteen}), the action difference between two fixed points is expressed as the algebraic area of the fundamental loop structures.  Therefore for homoclinic tangles in which the two fixed points coincide, the algebraic area of the fundamental loop is necessarily zero, ensuring zero difference in the actions of the fixed points.  This leads to the ``8" shaped turnstile structure in the simplest case, or structures with more crossings between the stable and unstable manifolds in order to maintain a zero algebraic area.   For heteroclinic tangles in which the two fixed points have different action functions, the algebraic area of the fundamental loop cannot vanish.  For the kicked rotor, this circle shaped structure is already demonstrated by Fig.~\ref{fig:twelve}.  

The topology of the loop in Fig.~\ref{fig:twelve} can be better illustrated if we shorten the unstable segment by iterating the loop inversely for one iteration, as shown in the lower panel of Fig.~\ref{fig:thirteen}.  The length of the stable and unstable segments are now closer to each other after the iteration, and the single loop is equivalent to a circle.  The upper panel of Fig.~\ref{fig:thirteen} is the well known turnstile structure of the homoclinic tangle formed by the upper branches of the unstable manifold of $(-0.5,0)$ and stable manifold of $(0.5,0)$, which is in clear contrast to the lower panel.  

\section{Conclusions}
\label{Conclusions}

There are many contexts in which one might be interested in the actions of periodic, heteroclinic, or homoclinic orbits.  It is worth noting that in the context of the trace formulae or semiclassical propagation of wave packets for chaotic systems, accurate calculation of periodic, heteroclinic or similarly homoclinic orbits is necessary for getting delicate quantum interference phenomena right.  A simple method to calculate such orbits in a strongly chaotic system is given here.  It works far better than any attempt to follow such a trajectory forward and inversely in time directly using the equations of motion.  The method relies on the structural stability of the invariant stable and unstable manifolds, and the exponential convergence of phase points in their neighborhoods towards them (unstable manifold forward in time, stable manifold inversely in time).  The high degree of accuracy of the method was demonstrated with the help of a stability analysis of the limiting phase points $(X_\alpha,X_\beta)$ as well as an exact relation between certain classical action differences and their equivalence to the related phase space areas.

An extremely interesting relation follows from the connection between phase space areas and classical action differences.  For the heteroclinic case, assuming a generic case in which the two fixed points have different actions as in the example shown, the closed path along the fundamental heteroclinic loop structure must enclose a non-vanishing area.  Remarkably, the intersections of the stable and unstable manifold has periodic orbit action differences built into it.  Contrast this to the homoclinic case where the loop structure is a turnstile.  There the closed path winds around in a figure eight path and the total enclosed area must vanish because $X_\alpha$ and $X_\beta$ belong to the same orbit and their action difference vanishes.  A heteroclinic loop structure cannot be a turnstile whose closed loop path integral vanishes.

The demonstration was carried out for a paradigm of chaos studies, the kicked rotor.  It takes the form of a discrete time dynamical map, but the same method could equally well be applied to continuous time dynamical systems by just creating an appropriate Poincare map.  Higher dimensional generalizations of the method is direct by using more than two points each for the interpolation of local unstable and stable manifolds.  For example,  in the case of a two-dimensional stable manifold intersecting a two-dimensional unstable manifold in four-dimensional phase space,  the local shape of the manifolds are interpolated by four points on the tangent plane,  so intersection can be located by intersecting the tangent planes.  Better accuracy is obtained by using denser set of points on the manifolds or more sophisticated polynomial interpolation techniques.   The applicability is therefore rather general.

\acknowledgments

The authors gratefully acknowledge that a general form of Eq.~\eqref{eq:eighteen} (which relates the action difference of two periodic orbits with the loop structure of their heteroclinic tangle) was derived jointly with Akira Shudo, Hiromitsu Harada, Kensuke Yoshida and one of the authors (JL) during a productive visit to Tokyo Metropolitan University and also gratefully acknowledge support for the travel.  

\bibliography{quantumchaos,classicalchaos,molecular,rmtmodify,general_ref}

\end{document}